# A Mobile Transient Internet Architecture

Henry N Jerez
Senior Research Scientist
Corporation for National Research Initiatives CNRI
Reston , Virginia 20191
Email: hjerez@cnri.reston.va.us

Joud Khoury and Chaouki Abdallah
Electrical and Computer Engineering Department
University of New Mexico
Albuquerque, NM 87131
Email: chaouki@ece.unm.edu / jkhoury@ece.unm.edu

*Abstract*— This paper describes a new architecture for transient mobile networks destined to merge existing and future network architectures, communication implementations and protocol operations by introducing a new paradigm to data delivery and identification. The main goal of our research is to enable seamless end-to-end communication between mobile and stationary devices across multiple networks and through multiple communication environments. The architecture establishes a set of infrastructure components and protocols that set the ground for a Persistent Identification Network (PIN). The basis for the operation of PIN is an identification space consisting of unique location independent identifiers similar to the ones implemented in the Handle system. Persistent Identifiers are used to identify and locate Digital Entities which can include devices, services, users and even traffic. The architecture establishes a primary connection independent logical structure that can operate over conventional networks or more advanced peer-to-peer aggregation networks. Communication is based on routing pools and novel protocols for routing data across several abstraction levels of the network, regardless of the end-points' current association and state. The architecture also postulates a new type of network referred to as the *Green Network*. The *Green Network* has protocols to coordinate routing traffic and to allow for the identification and authentication of devices, services , users and content characterized as Digital Entities. Transmission is assumed to initiate and terminate at transient physical locations. The network implements every reasonable effort to coordinate a prompt delivery to the transient end-points using what ever means available. This paper is a conceptual logical model of the intended architecture and specifics about its particular components and their implementations will be discussed in future papers.

## I. Introduction

There are two main issues that can be identified regarding the formal structure of the network over which mobile and transient devices must currently operate:

1) The current implementation of the Internet is based on location and association aware communication that in effect resembles a virtual circuit path communication. The problem is that in high latency, variable and mobile networks both ends of the circuit are frequently changing which in turn results in extensive retransmission and delays as well as communication failures.
2) The current network communication schema links the identification of the communication devices to a particular location at the physical level by means of the IP association and distribution. Real migration and mobility are therefore hard to fit in a rigid primarily static hierarchical structure.
3) The current internet does not natively provide any off-line communication mechanisms for transient users moving across multiple networks with intermediate disconnections.

We propose a new paradigm to transient communication networks. We essentially contend that mobile and transient devices should be part of a "Green Network", a transient mobile architecture that treats nodes and traffic as Digital Entities. The Digital Entities are identified with unique and persistent identifiers. Traffic is treated as data pools being exchanged between distinct entities by means of a replicating self propagating self adjusting network. The Green Network effectively isolates the data exchange and delivery from the communicating parties. Rather than having the end-points coordinate and adjust the communication by retransmitting and controlling the flow of their traffic; these nodes are now part of a network of dual purpose entities uniquely identified with location independent identifiers, that produce and route pools of data (pods). These PODs are in turn identified by persistent identifiers as well . Once a certain POD has been deposited into the Green Network, it (the network) will assume the responsibility of routing the traffic to fit the end nodes movement given the current characteristics of the network. Data is not moved towards a particular destination but rather routed into the Area of Influence of a particular device. The Area of Influence (AoI) is expressed in terms of the general area of communication that a particular ad-hoc node is known to be associated with. This way a roaming node will have the data delivered to it through its current AoI. This paper reports on current status of our work, describes the full scope of the architecture and outlines our future research path to complete a fully distributed implementation of this architecture. Areas of Influence are composed by a set of nodes that form an ever expanding and growing network that merges several layers and levels. These nodes and the AoIs they form are part of the network core and edge. They are in fact capable to expand this edge seamlessly to incorporate even more nodes into the network.

The overall Mobile Transient Network builds on the original logical model of the internet to form a logical network that allows the effective merging of heterogeneous networks without forcing them to modify their communication protocol but rather their logical coordination mechanism.

## II. GENERAL CONSIDERATIONS

In order to foster inter-operability and seamless interaction we propose the use of a universal persistent identifier that is location and association independent. The Handle System [17], a globally distributed persistent identification system, has most of the required characteristics in terms of security, reliability and scalability needed to identify not only content and devices but also traffic packets and even users. We have successfully used this system to implement the early stages and the first test bed of some concepts discussed in this paper. The full realization of this architecture nevertheless, calls for certain features that may require an extension or even a re-implementation of the current Handle System. These features and the proposed characteristics of this Distributed Persistent Identification Network are Discussed in the Ongoing and Future work section.

## III. MOBILITY, PERSISTENCE AND AGENTS

Mobility is traditionally associated with physical network address association and re-association. As we mentioned before this means that communications are based in the concept that a full virtual circuit between one end and the other is, at least during the course of a particular communication exchange, immutable. Therefore communication is expected to occur within a set of formally identified and immutable devices associated with particular instantaneous connections. As a result the communication is static in nature, especially because the routing mechanisms used by the current internet implementation are based on network level addressing. This communication additionally depends on the ability of a particular node to implement the communication protocol of the initiator and to be physically associated with an address (typically IP) in the currently deployed network. We propose a different type of communication that is persistent in nature and oblivious to network address volatility. The persistence that we describe is the result of a higher level indirection based resolution mechanism that uses persistent identifiers as in [1], [18]. True persistence is therefore the result of an independent network identification mechanism that abstractly identifies each device and piece of data being transmitted regardless of its communication and interaction mechanisms. The identification system that we propose is essentially invariant across time and network association. This translates into communications that can survive not only network re-association and migration but also physical disconnections and relocations.

Persistence per-se requires the presence of independent actuators that perform maintenance tasks and implement the overall architecture policies. We propose a series of agents that are destined to perform these tasks. Such agents are responsible for updating and disseminating information inside the system. This information handling is destined to update the global persistent identifiers and establish logical coordination. The Agents we propose are based on the concepts introduced in [13] about Knowbots [2] which are mobile agents that are addressable as long as they are associated with a particular service station. Our agents characterization; however, is different in the sense that we conceptualize our communicating parties as agents. These agents are capable of interfacing directly with the operating system and associated hardware of particular hosting nodes. They are addressable regardless of their particular association through persistent identifiers; which enables them to be part of a set of flexible overlay networks. The globally signed and authenticated agents are capable of updating crucial global information and routing traffic based on this logically persistent infrastructure. Hence, persistence, which was traditionally adversely affected by mobility, is achieved not only trough the existence of a logical persistent infrastructure, but also through a set of agents that contribute to the general system freshness and stability. A conceptualization of this network is shown in Figure 1.

## IV. DEVICES AND SERVICES

In order to guarantee the stability and effectiveness of the persistent network, global and persistent indirection identifiers are used to abstract device and service identification from their particular network association and location. The Handle System [1] allows us to use a common infrastructure to identify particular devices and their services as well as the data flowing through them. Particular devices will map locally and in the context of their current means of communication to a particular network and physical address. At the same time, services will map locally to particular identifiers possibly inside the particular operating system and environment in which they are set to exist. Both types of addresses are bound to change over time and are therefore prime candidates for persistent identification using a set of Handle identifiers to abstract their locality transforming them into globally addressable resources. The maintenance and implementation of the persistent identifiers at the level of the devices and their services is traditionally handled by agents that reside inside the moving devices. There also exists a set of devices which are either incapable or unwilling to hold diverse or flexible enough agents to correctly implement the logical network structure. This is the case of devices with very limited hardware resources such as sensors, embedded systems and some application specific hardware devices like Voice-over-IP phones. In order to enable these devices to participate in the persistently identified network we introduce Persistent Coordinated Translation (PCT) gateways. These gateways set the framework for the dynamic incorporation of multiple devices to the overall inter network persistent overlay. They not only assume delegated agent functions for the devices but also provide these agents with the necessary hardware resources that would enable them to communicate and expand their associated device connectivity features while implementing the global overall architecture.

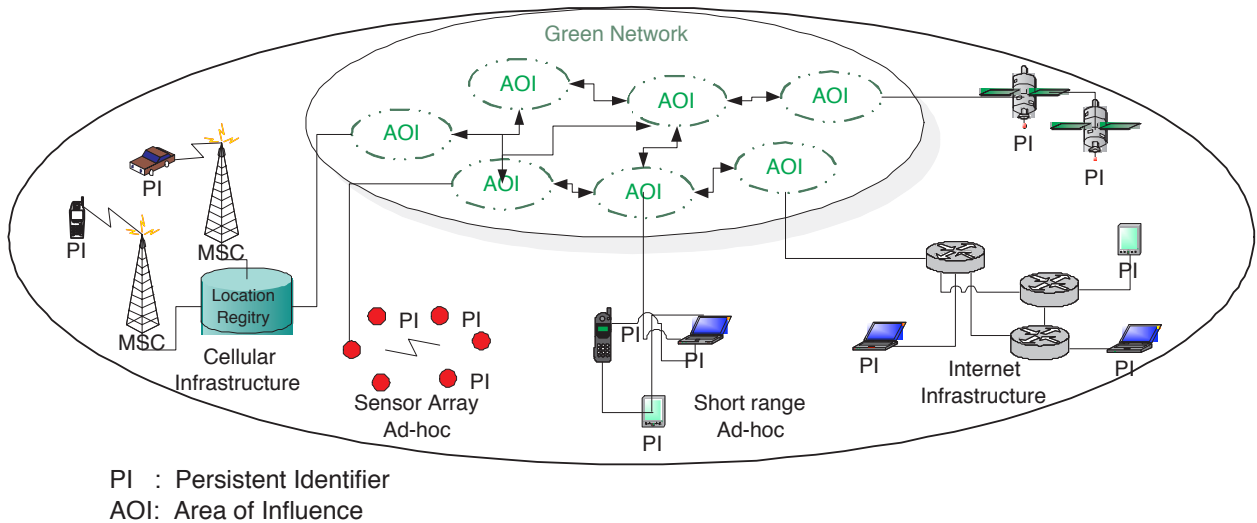

Fig. 1. General Architecture Conceptualization

## V. AREAS OF INFLUENCE

Before we discuss the translation gateway concept, we characterize a novel postulated set of ad-hoc networks. We shall call these networks *Areas of Influence* (AoI). An AoI is composed of a defined set of entities that communicate with each other by means of a common protocol. These AOIs can aggregate themselves into larger AoIs through the implementation of a layered approach performing aggregation and delegation operations.

In terms of the traffic and data propagation, the components of traditional ad-hoc networks can only communicate natively with the nodes that they can directly interact with. ((This translates into all the devices that implement a common communication protocol with the device and are within the nodes reach.)) This concept is where the definition of an actual Area of Influence and its actual name originate. We contend that a structured network is a relatively special case of an ad-hoc network. Therefore, an architectural definition for ad-hoc networks could apply to structured communication networks as well. Based on this assumption, *we define an area of influence as a local ad-hoc based communication community that defines its own communication protocols and network architecture implementations.*

Some instances of AoIs are:

- A set of nodes communicating in a structures network. These AoIs could be a set of nodes communicating behind an access point or more elaborated local, wide or metropolitan area networks.
- A set of nodes communicating using a common cellular protocol. Examples of AoIs in a GSM environment could include the nodes belonging to a cell and communicating with a particular Base Station or even a complete cluster of cells that could be perceived as an AoI.
- A set of nodes communicating using a common ad-hoc protocol. Some examples are sensor networks and wireless personal area networks.

Figure 2 illustrates some of these network instances and their characterization as Areas of Influence AOIs. These AoIs communicate with each other by means of translation gateways such as Base Stations (BS) Switching Centers (MSC) in cellular networks, the wireless Access Points (AP) connected to the local area networks in the case of WiFi networks and the routers in the case of LANs. The gateways, identified with persistent identifiers, are traditionally conceived as protocol translation gateways. Nonetheless, they can provide more sophisticated services including, for instance, the resolution of persistent identifiers and the conformation of overlay networks. These overlay networks enable the distribution of persistent identification data and current persistent identifier and area of influence association. These services are what makes the gateways different in our architecture. We refer to the gateways as Persistent Coordinated Translation Gateways or PCT-Gateways due to the fact that they are also identified persistently in the architecture.

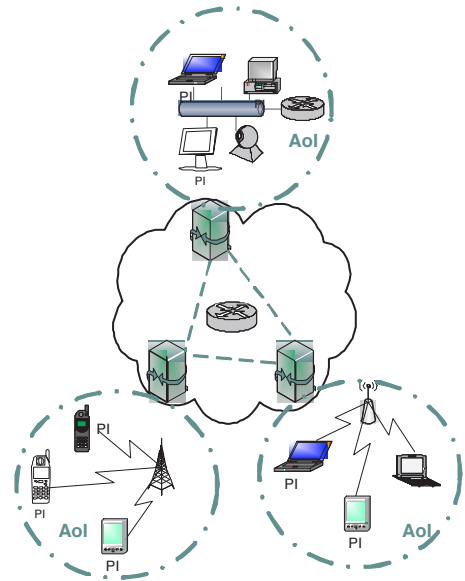

Fig. 2. Multiple Areas of Influence *AoI*s interacting with each other

## VI. PCT-GATEWAYS AND OPEN DEVICE ACCESS PROTOCOL

Throughout the years, there has been a push to create an Internet that puts the least amount of intelligence and configurability in the underlying network infrastructure and push any intelligence to the edges and termination of the network obeying the *end-to-end argument* [15]. We postulate a system in which a third intermediate level that prolongs the actual reach of the Internet is introduced. We introduce an expansion layer that appears to the internet as an edge service as well as an infrastructure component. The expansion layer uses the edge nodes to provide expansion services that turn them into basic communication infrastructure components. This layer and its components relieve roaming and mobile devices as well as heterogeneous networks from the burdens of compatibility and constrained implementations of larger protocols and non native stacks. The expansion layer is depicted in Figure 3.

The Persistent Coordinated Translation (PCT) gateway is intended, in essence, to provide a seamless implementation of an overlay persistent network to diverse often non standard network implementations. It allows these networks to communicate not only with the overall backbone infrastructure, but also amongst themselves. The PCT gateway is responsible for empowering persistent identifier association and network protocol translation to homogenize the communication space amongst heterogeneous networks. PCT gateways may lay in the midst of logical network junctions or at the edges of homogenous network connectivity structures to expand and reshape their reach onto other networks and devices. We identify two types of PCT Gateways:

- Edge PCT Gateways or E-PCT Gateways: Edge PCT gateways implement a novel protocol called the Open Device access protocol postulated by Jerez in [6]. The protocol is destined to discover and associate heterogeneous devices and the networks through them with the overall persistent infrastructure. These gateways serve two types of devices:
    – Devices that are capable of maintaining their own agents but need intermediate deployment and staging servers mainly due to resource constraints.
    – Devices that are not capable of deploying an actual agent but carry within them the means to authenticate and validate with a particular surrogate agent.

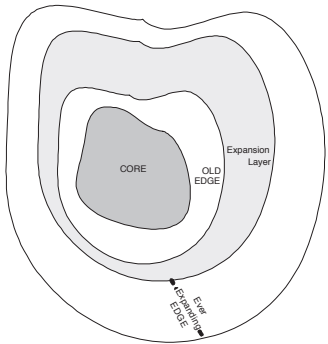

Fig. 3. Intermediate layer mediates between core and edge layers to add more flexibility.

  These gateways are responsible for associating and updating device and service handle identifiers while providing basic network translation resources. E-PCT gateways can provide intermediate or staging bases for agents to deploy and implement services with their associated devices, which they discover and associate with the overall architecture. These agents and the E-PCT gateway in turn perform physical, network and software level translation operations as well as regular overlay network functions destined to maintain the persistent identifier consistency. The E-PCT gateways also provide multiple AoI inter connectivity and routing services as we shall see in later sections. When E-PCT gateways operate as AoI intermediaries they are responsible for routing traffic between AoIs and assume an information broker role in the architecture. It is important to notice that any node with the accurate resources could turn into an E-PCT gateway.

- Interface PCT Gateways or I-PCT Gateways: Interface PCT gateways or I-PCTs handle the migration of non persistent instantaneous traffic to transient persistent networks and vice versa. They are also responsible for interfacing current technologies with the persistent resources and facilities of our transient infrastructure. The agents residing in these gateways perform on the fly protocol translations such as DNS-to-Handle and vice versa as well as application specific implementations. This is precisely the type of gateway used by Khoury [8] for integration of DNS infrastructure with the Voice-over-IP roaming infrastructure. The I-PCT gateways provide application level translation and service integration rather than formal traffic routing.

The effective combination of E-PCT and I-PCT Gateways enables the merging and adaptation of the basic transient network infrastructure behind the new routing protocols and content distribution in our architecture. The resulting network is a basic bandwidth aggregation and propagation system that builds on of top persistence and indirection mechanisms to pursue a mixed routing, dynamic and self organizing network. This bandwidth aggregation characteristics are discussed in the Ongoing Work data routing section.

## VII. TRANSIENT MOBILE NETWORK, A HYBRID NETWORK

We have mentioned several times the transient mobile network. A transient mobile network is a network implementation that supports mobility, persistence and pervasiveness by design assuming all connections between its entities to be transient. These entities may communicate in a variety of ways to implement either traditional communication mechanisms or a modified approach to swarming networks [12] that we refer to as a Green Networks. As we mentioned before Green Networks handle routing and content transfer based on persistent identifiers and Areas of Influence. This characteristics are therefor reflected in the transient network as well.

The transient network in a structured peer-to-peer fashion forms a distributed network divided into areas of influence that build on top of persistent identifier based routing protocols. The overlay assumes the network as a collection of agents communicating and transferring information amongst themselves. The different overlay networks that are implemented on top of this vanilla architecture are capable of reconfiguring and coordinating traffic and routes depending on inline data used to shape and control traffic. Routing and transfer are therefore contingent on global parameters and on a series of principles that guide the overall transient network architecture. Since actual routing and protocol coordination for these networks depends on the overall policies implemented by the agents, a multitude of overlay networks may be deployed on top of these networks. Traffic is intended and assumed to initiate and terminate at transient locations identified by persistent identifiers such as Handles [16].

Even though the PCT gateway components would thrive in a green network environment; they are in fact independent enough to be deployed as interfaces for regular internetworking standards with traditional devices as well as different green network implementations. In fact, PCT gateways are capable of forming logical green networks amongst themselves. This translates into seamless integration of heterogeneous networks into a larger more effective Hybrid network. Green networks are designed to accommodate all of the three types of Networks reflected in Figure 4:

- Structured Static: Networks with fairly constant and reliable nodes in which changes are unlikely or at least predictable.
- Structured Mobile : Networks of Mobile devices that are associated with structured access points, such as wireless hot spots and cellular networks.
- Unstructured ad-hoc networks: Networks formed on the fly by sensor and radio networks, ever changing and unpredictable in nature.

We can therefore characterize the transient mobile network as a Hybrid network by nature that merges transient, mobile and traditional networks. The unifying technology across all these different types of networks is the persistent identifier and indirection technology. The persistent identifiers that are implemented as a global sectionable name space, effectively distribute and merge the prefix name space across all AoIs and component nodes. This hybrid network not only allows the co-existence of transient and conventional networks but, in some measure, merges them allowing our overlay architecture to be truly global.

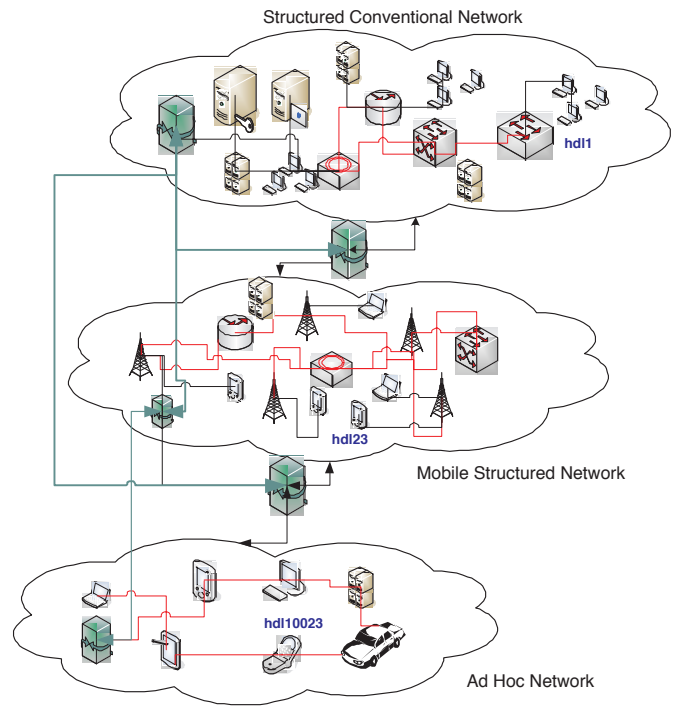

Fig. 4. Hybrid Network Characterization

## VIII. STRUCTURED TRANSIENT NETWORKS VS SWARMING TRANSIENT NETWORKS

Transient network implementations can follow either a traditional fully overlay approach or a more progressive swarming peer-to-peer based approach. In the fully overlay based Transient network approach, pre-existing infrastructure that provides a secure and persistent identification network, such as the Handle System [1], is assumed to exist. With this approach, PCT Gateways behave mostly as I-PCTs providing service and basic protocol translation without much additional routing services. This is the approach that we have used so far to implement persistent indirection for our mobile nodes. We have primarily replaced the domain name or IP-to-device association with a more flexible and abstract persistent identifier-to-Digital Entity [1] association. This presents a fresh approach to mobility by allowing the Digital Entities to be formally identifiable and addressable regardless of their current association. More details about our previous works with this type of transient network is presented in the next section.

In addition to the more traditional purely overly based approach, our conceptualization of mobile transient networks allows for a more independent and flexible implementation. The concepts of ad-hoc networking, swarming technologies as well as peer-to-peer based routing, enable this approach. Unlike the traditional or purely overlay based approach, the swarming type of transient networks that we more formally characterize as a Green Network, does not assume a pre-exiting network. It does not assume a set of already deployed networks and paths.

[1]Digital Entities here include devices, services or users

It enables a grass roots approach to network conformation organization and communication. This is achieved by means of a series of conformation and routing rules that help it realize natively a complete set of new services and applications. Green networks extensively use E-PCT Gateways, peer-to-peer and swarming technology notions to propagate and replicate data in order to generate aggregated traffic and communication rather than single path data transfers. Green Networks exploit the AoI's independence and flexibility while relying on the persistent identifier network to provide inter-operability. We will discuss the proposed characteristics of these networks, currently under research, in section IX.

### A. Structured transient network previous work and results

We have successfully implemented some I-PCT and E-PCT gateways that have enabled us to deliver user roaming and mobility across multiple networks and environments. As part of our work, we have specifically provided DNS-to-HANDLE translation that enables our systems not only to expedite DNS resolution, but also to extend its functionality. Khoury [9] has already shown how this allows for basic access from regular IP enabled devices to non-IP devices identified by handles [11] and also ensures compatibility with existing applications that exploit DNS addressing. Most of our work currently uses the Handle System [1] as its main source for persistent identifier and resource association. The PCT gateways involved in the communication process allow us to interact with currently deployed environment by either substituting or alternating the identification resolution with the currently deployed DNS system. Since we rely on the readily deployed PKI enabled Handle System, our performance in terms of resolution is either comparable to or better than that of the traditional resolutions of existing services as shown in Figure 5.

The work presented by Khoury [9], [8] has demonstrated that even the sole use of the Handle System along with the concepts we have introduced so far can improve current mobility applications and deliver new services. There is nevertheless a fundamental limitation in traditional overlay applications which is that of the underlying communication mechanism. The current infrastructure addresses all traffic based on IP addresses or their DNS entries which are rather static and take a long time to propagate updates. The final realization of our architecture is only possible with the correct deployment of more advanced E-PCT gateways that enable a different approach to routing and propagation. This approach could probably substitute or extend IP in the future but this is not a requirement at this point.

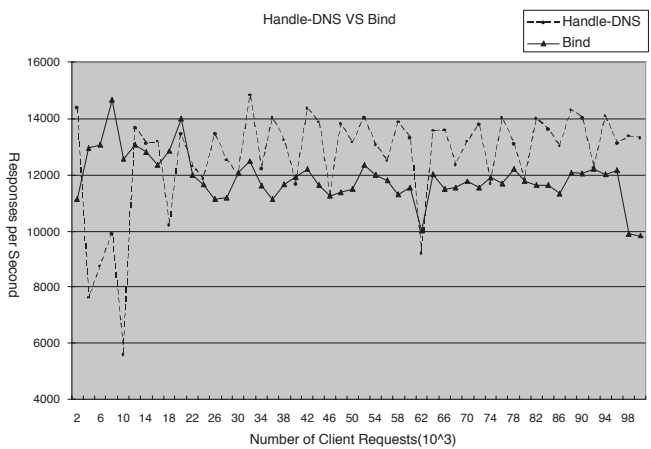

Fig. 5. *Handle* load performance measurement as of August 2005. Obtained through the courtesy of Mr. Sam Sun and CN-NIC.

## IX. Ongoing work with transient swarming networks: The Green Network

Transient swarming networks are the basis for the proposed Green Network topology that complements the transient network architecture. These networks are based on the concept that a series of Areas of Influence will interact with other AoIs to form larger networks. Each AoI implements its own local addressing mechanisms, communication protocols and routing algorithms. The resulting network is an aggregation of AoIs that intercommunicate through E-PCT Gateways that, in turn, route traffic between AoIs. These gateways base their routing decisions on queries to a coordination network that stores the persistent identifiers-to-AoI associations. Hence, each AoI is solely concerned with propagating the traffic from a particular constituent end-point to the next AoI. Global coordination exists but is handled at the AoI level rather than the individual node, reducing overhead and streamlining communications.

Green Networks assume a clean slate approach to network conformation based on ad-hoc network principles and are capable of reassuming this behavior at any point in time to guarantee resilience. This means that Green Networks are intended to rearrange to their current characteristics automatically. This results in the ability of green networks that become isolated due to massive network failures to automatically seek to re-balance and re-constitute coherent networks.

Once a green network is implemented, AoIs can become either transient reorganizing structures in the case of ad-hoc networks, or highly structured and stable architectures in the case of traditional infrastructure networks. Green networks embrace this characteristic to prevent network ossification and provide a fertile ground for innovation and adaptation.

### A. Network conformation

The Green Network is formed by the incorporation of peer nodes that start their original network integration by scanning their surroundings and initiating a handshaking protocol in order to discover other nodes and their particular AoIs. When two nodes that are not associated with an active AoI find each other, they mutually challenge to determine a new AoI. Nodes may be part of more than one AoI, in which case, they must declare a primary AoI. An intrinsic secure communication is expected out of the nodes that conform to a particular Area of Influence. This security is the result of both the self computation of particular individually signed identifiers, and the delegation of identification name spaces to avoid duplication and guarantee identifier verification. The identifiers themselves are intended to be handled by a Distributed Persistent Identification Network (D-PIN). This network is responsible for storing and resolving the persistent identifier attributes and is expected to be highly secure, robust and self organizing thanks to the extensive use of secure authentication and communication ecnryption.

So far, applications of our previous work have used the IPv6/IPv4 protocols to confer information at the network level. But logically, in the Green Network, the unit of identification is the Persistent Identifier. Persistent Identifiers in their Handle [18] form are already used as content identifiers [5], device and service identifiers [9] and even user identifiers [8]. In the Green Networks, we propose that persistent identifiers be used to track transmitted data as well . The key to seamless migration, mobility and integration not only of entities, devices and content but possibly even of transmission data, is the actual persistence of these identifiers and their distributed resolution. Communications may then be perceived as data moving amongst AoIs implementing totally different protocols while maintaining its pool of data (pod) structure and identifying itself with persistent identifiers.

The present implementation of the Handle System [17], currently the largest distributed persistent identification network, uses a schema in which all non-cached lookup queries must go through a global registry that locates a local service responsible for resolving the respective handle. If the user knows the address of the local service, the step of using the global registry can be avoided, but this is not the general case. The need for a global registry can be avoided by organizing the persistent identifier servers into an overlay thus forming what we call a Distributed Persistent Identification Network (D-PIN). As part of this overlay, queries are propagated amongst the composing nodes of a structured P2P network. With D-PIN, the now centralized functions of administration and resolution of the Handle System become distributed.

*1) Distributed Persistent Identifier Network: D-PIN:* The proposed Distributed Persistent Identification Network is a structured p2p network which allows a delegation architecture for its global name space. The delegation of the name space is directly related to specific AoIs that comprise the transient network. These AoIs are responsible for implementing and maintaining a p2p based distributed identification mechanism. The D-PIN network is thus composed by several Areas of Influence and delegated portions of it are maintained for all the members of a particular Area of Influence. Figure 6 illustrates the interaction between multiple AoIs and their Persistent Identifier Networks. When different AoIs interact, they not only establish the PCT gateways responsible for their interaction but also merge and load balance their PINs. This merging of PINs is possible due to the delegated architecture of the D-PIN name space.

Upon successful interaction, the PIN data itself is load balanced and replicated to guarantee pervasiveness and robustness. The resolution network is therefore a Distributed PIN as shown in figure 6.

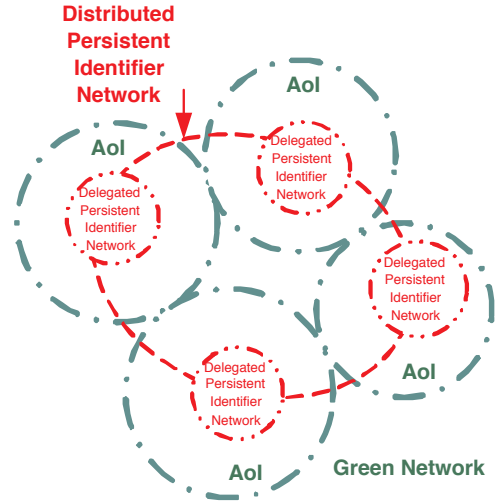

Fig. 6. Interaction of AoI and Persistent Identification Network

In the event of a catastrophic failure or isolation, the delegated PIN assures local PIN resolution and returns to an isolated mode of operation recalculating its own name space associations based on the presence of new roaming clients. Persistent Identifiers are permanent and pervasive. PIN validation occurs upon the arrival of device association or re-association requests to a particular AOI. In the event that the PIN is working in an isolated mode, the new clients are flagged for validation upon AoI interaction. PIN resolution messages indicate Persistent Identifier validity at the time of resolution to avoid impersonations.

The D-PIN architecture itself is secured at several levels , from the creation of unique identifiers to the certification and encryption of stored data. The communications that take place amongst the nodes may also be encrypted. This way identity may not only be securely validated at every instance but also administered and propagated securely. This provides a basic block of trust provided by agents secure and validated identification during routing.

*B. Data Routing*

As we mentioned earlier in this paper, information transferred inside the swarming version of the Green Network is characterized as Pools of Data or pods. These pods are components of larger files identified themselves with persistent identifiers [14]. Each pool of data is routed and moved along depending on a set of routing considerations based on content priority as well as overall network coordination. The green network itself has two different types of traffic that mimic somehow TCP /IP:

- A control and coordination traffic
- A payload traffic

The control and coordination traffic is destined to both semi-static agents residing in what are considered infrastructure nodes(this characteristic may change at any point in time) and entities collocated or associated with a particular node. From a routing perspective, each node is defined as a hosting site for one or more agents. This holds true in all instances except that of nodes that are limited by their hardware characteristics and are not able to hold a valid agent i.e sensors. In these cases, the nodes must be able to hold a valid unique mean of identification that can be used to spawn a valid agent for them.

The inter-agent communication accounts for the actual network traffic. The global coordination and the routing considerations are based on information dissemination from both types of agents (infrastructure and mobile ones). Particular routing decisions are taken based on the cumulus of information present in a particular AoI at any particular point in time. AoIs, PCT gateways and therefore agents are inherently selfish, they route and replicate pods according to two premises:

- First, follow a set of priorities.
- Second, get pools of data across to the next area of influence.

A sketch of data being routed between two nodes is depicted in Figure 7. Traffic originating from node *PI2* towards node *PI234* is always pushed closer to the latter's Area of Influence. Routing is based on persistent identifiers and is independent of the end node's association and state. The architecture will allow for traffic to reach *PI234*, even if the latter relocates to another AoI. This characteristic is the direct result of routing based on persistent identifiers. The classification of backbone or trunking nodes is not a function of manual configuration but rather an automatic computation that is the result of experience, coordination and adaptation. Much like a neural network, measurements of reliability and well known paths will result in certain communication routes becoming more common. This however does not compromise stability or scalability since additional routes are tried and created upon certain conditions in the network being met. This task follows basic load balance and traffic distribution in the system. The previous statement brings along the discussion of survivability and robustness. It is clear that the more nodes are present cooperating with transmission (the level at which any particular node may cooperate with the transmission is related to a set of hardware, security and environment considerations), the faster and more reliable the transmission will be.

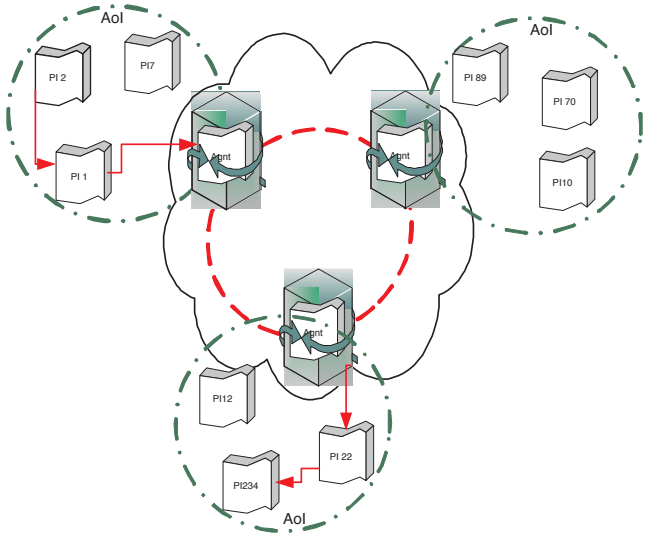

Fig. 7. sketch of Inter-AoI routing

## C. Data Propagation

As multiple nodes are communicating information in the form of pods; data is being replicated across nodes to guarantee persistence and aggregated bandwidth. How much data is replicated and what its replication and transmission priority is, depends on priority measurements associated with the pods and their originating and terminating agents. These pools are earmarked with particular globally accepted priorities and the agents residing in the nodes are capable of implementing decisions and logic based on globally shared Green Network routing rules. This information may be updated at any point in time by globally signed roaming agents that deliver executable code or instructions. The mobile agents are constantly moving and updating the routing considerations on the PCT gateways. Policies are intended to be in place to have global network reconfigurations initiated from any point in the network upon request.

As with any network, propagation plays a key role. The system is not intended to have any centralized features which in turn means that it is robust and distributed. Propagation translates into a swarming type of connection. This connections provides either multi cast communication with pure replication of pods or an aggregated transmission bandwidth similar to that of torrent and onion networks [3], [10] as depicted in figure 8. The network is also robust enough to automatically reshape is distribution based on demand, mimicking the Akamai effect [4].

Green networks also provide for the existence of special purpose agents that effectively behave as self contained repositories for transient data in the network as postulated by Kahn in [7]. This in turn allows communication to nodes that are operating in a disconnected mode. In this approach, nodes will appear, drop information in the network, receive information and then disconnect. A whole set of priorities can be associated with this type of traffic that allows for the computation of optimal paths and established trunking paths. The formal rules for data propagation inside Green Networks can be defined as an evolution of the routing rules:

1) Respect and obey global policies.
2) Respect and obey internal characteristics and conditions.
3) Get pools of data through to the next Area of Influence.

Based on these characteristics; failures or outages in the systems will be reflected on the least priority traffic first and the most important coordination traffic last. Increases and benefits unless otherwise stipulated will always benefit priority traffic first also. The more powerful the communication mechanisms are, or the more nodes present, the better the network experience will be for everyone. This translates into a network of high efficiency for crucial traffic.

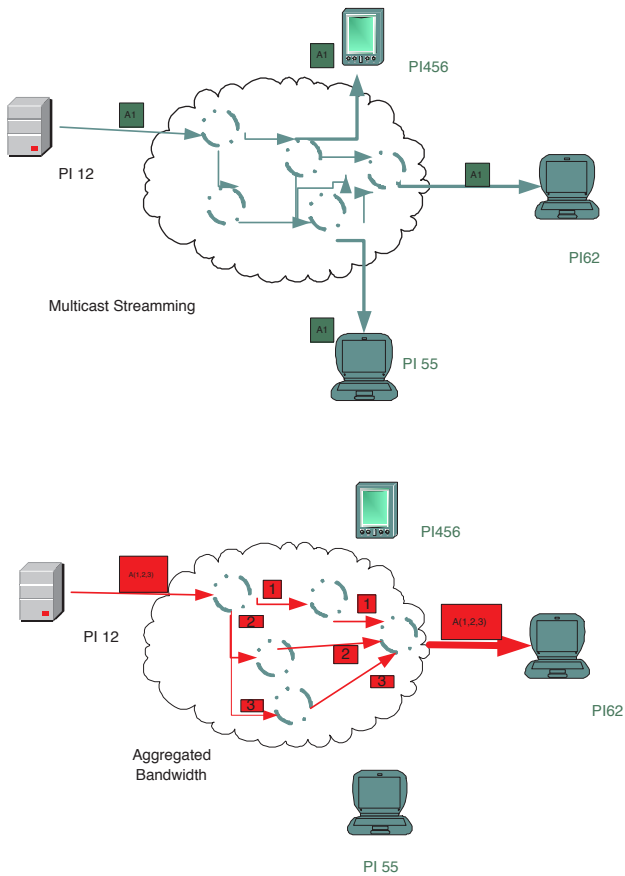

Fig. 8. MUlticast and aggregated bandwidth communication with pods

### D. Management, Deployment and Conceptualization

Throughout this discussion we have skimmed at the formal representation of the nodes themselves. In fact we use an evolution of the Digital Object Architecture postulated initially by Kahn in [13]. The Digital Object Architecture was originally developed by the Corporation for National Research Initiatives (CNRI) to re-conceptualize the Internet, based on managing information rather than simply communicating information from one computer to another. Its key attributes are a common currency based on "digital objects", a resolution mechanism that maps identifiers for digital objects into state information about the objects, repositories which may be static or mobile, and metadata registries that may be used to retain information about the objects for purposes of search and data mining.

In our architecture, each communicating resource is mapped into a digital abstraction, which forms a digital object that is uniquely identified by a persistent identifier. Such digital objects are considered persistent and independent of their current physical and geographic attributes. This abstraction allows persistent addressing and communication with these entities regardless of their current association, location or means of communication. For example, such an entity could move seamlessly from one network environment to another and be seamlessly incorporated, provided it met the administrative requirements. Besides, the abstraction allows for information at any level to be uniformly addressable and accessible at any other level, subject only to administrative constraints. Digital Entities may be network end-points, network components, users, applications, agents and backbone building blocks all of which exploit the digital object architecture to seamlessly enable pervasive, ubiquitous communications and system implementations.

The inherent secure distributed administration, management and coordination of this network is assured through the implementation of the previously mentioned Distributed Persistent Identification Network. The overall transient mobile network architecture is conceptualized as a logical structure whose implementation is isolated from its components while administering, designing and relocating them from a particular hardware instantiation to another. By abstracting the component descriptions from the specific network implementation, and providing them with a common set of expected behaviors and interfaces, it provides local implementers with relatively unconstrained flexibility to develop, manage and expand the network implementation. Figure 9 provides an image of the conceptual view that network administrators are presented with.

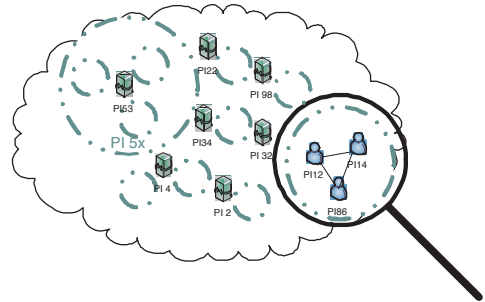

Fig. 9. Network administrator new conceptual view

## X. CONCLUSIONS AND FUTURE WORK

We have presented in this paper the basic conceptualization of a mobile transient network and its components. This is a heterogeneous network architecture based on persistently identified abstractions which map, at any point in time, to their current implementation, location, and other attributes required for network communication. We refer to this as the transient mobile network architecture. Each resource on the network, including network end-points, network components, users, applications, and backbone building blocks, is abstracted as a digital object and assigned a unique identifier within a distributed persistent identification system. It is that identification system which provides the mapping of an abstraction to the current state of the resource. This architecture begins by assuming a mobile world with transient devices communicating across a fully distributed environment. This characterization not only enables a native answer to the requirements of these devices that represent the fastest growing set of components for the internet but also provides a flexible structure to welcome the new devices, services and users of the future.

We are currently working on industry financed test beds of the digital object characterization for administration of backbone building blocks and a brand new distributed persistent identification network. This work should allow us to expand our application of the architecture and finally realize its potential as a breeding ground for new protocols and communication architectures.

ACKNOWLEDGMENT

The authors would like to acknowledge the contribution of several testing institutions and the support of both government and industry based sponsors of this research.